\input{aipcheck}

\documentclass[
    ,final            
]
  {aipproc}

\layoutstyle{6x9}

\begin{document}

\title{Models of flavour with discrete symmetries}

\classification{11.30.Hv, 12.15.Ff,14.60.Pq }
\keywords      {Flavor symmetries; Quark and lepton masses and mixings;
Neutrino masses and mixings }

\author{A. Mondrag\'on}{
  address={Instituto de    F\'{\i}sica, UNAM,
    Apdo. Postal 20-364,  01000 M{\'e}xico D.F., \ M{\'e}xico}
}



\begin{abstract}
  We briefly review some recent developments in theoretical
  models of fermion masses, mixings and CP violation with discrete
  non-Abelian symmetries. Then, we explain the main ideas of a
  recently proposed Minimal $S_{3}-$invariant Extension of the
  Standard Model and its application to a unified analysis of masses,
  mixings and CP violation in the leptonic and quark sectors as well
  as the explicit computation of the $V_{PMNS}$ and $V_{CKM}$ mixing
  matrices.
\end{abstract}

\maketitle

\section{Introduction}
In the last six or seven years, great advances have been made in the
experimental knowledge of flavour physics, fermion masses and mixings
and CP violation. These advances initiated a huge upsurge of
theoretical activity aimed at uncovering the nature of this new
physics. In the next section of this paper I will very briefly outline
some recent theoretical developments on models of flavour with
discrete non-Abelian symmetries in which the participation of the
mexican conmunity of particles and fields is visible. Section 3 is
devoted to a brief explanation of the recently proposed Minimal
$S_{3}-$invariant Extension of the Standard Model \cite{kubo1}. The
paper ends with a short summary and some conclusions.

\section{Models of flavour with discrete symmetries}
The history of models for the quark mass matrices may possibly be
traced back to Weinberg's\cite{weinberg} observation that the Gatto, Sartori,
Tonin\cite{gatto} relation for the Cabbibo angle may be expressed as a
relation between the Cabbibo angle and the quark masses of the first
two generations,
\begin{eqnarray}\label{vus}
V_{us} \approx \sqrt{\frac{m_{d}}{m_{s}},}
\end{eqnarray}
and that mass matrices of the form
\begin{eqnarray}\label{mu}
M_{u}=\pmatrix{
m_{u} & 0 \cr
0 & m_{c}
}, \hspace{1cm} 
M_{d} = \pmatrix{
0 & p \cr
p & q 
}
\end{eqnarray}
can account for the approximate equality (\ref{vus}). As a
consequence, in the early approaches to the problem of quark masses
and mixings, it was natural to postulate that some entries in the
Yukawa matrix were equal to zero, the so called ``texture zeroes''
\cite{Fritzsc1,Fritzsc3}, thereby reducing the number of free
parameters of the theory. Since then many approaches have been
developed in the context of different theoretical and phenomenological
models.

In the Standard $[SU_{C}(3)\times SU_{L}(2)\times U_{Y}(1)]$ Model of the
strong, weak and electromagnetic interactions, it is the Higgs
mechanism that provides a theoretically consistent framework to
generate masses for gauge bosons and fermions - the latter acquire
masses after spontaneous breaking of the $SU(2)$ gauge symmetry,
through the Yukawa couplings and the vacuum expectation value of the
neutral Higgs field. However, this framework can neither predict the
values of fermion masses nor interpret the observed hierarchy of their
spectra. Hence, the three charged lepton masses, the six quark masses
as well as the four parameters in the quark mixing matrix are free
parameters of the Standard Model. As a straightforward consequence of
the symmetry structure of the Standard Model, the renormalizable
Yukawa couplings do not allow neutrino masses, although they can be
introduced through the addition of non-renormalizable,
higher-dimensional operators, presumably originating in physics beyond
the Standard Model.

In the late 70's and early 80's, there already were a few promising
indications for theoretical structures beyond the Standard Model which
addressed the fermion mass problem. For simple symmetry breaking
schemes, grand unification can relate quark and lepton masses. The
most promissing of such relations is the equality $m_{b}=m_{\tau}$, a
result which applies at the Grand Unified Theory (GUT) scale.
Radiative corrections are dominated by QCD interactions which increase
the bottom quark mass in fair agrement with experiment. A fairly
straightforward supersymmetric generalization of such GUT relations,
which involves a new family symmetry, also provides the
Georgi-Jarlskog \cite{georgi} relations between the down quarks and charged
leptons of the first two generations.  Supersymmetry also enables the
gauge couplings to meet at the GUT scale to give a self-consistent
unification picture \cite{ellis,amaldi,giunti,langacker}.

The past ten years have seen great advances in the experimental and
theoretical knowledge of flavour physics, CP-violation and fermion
masses. In 1999, direct CP-violation in the Kaon system was
established through the NA48 (CERN) and KTeV(FNAL) collaborations. In
this decade, huge experimental efforts have been made to further
explore CP-violation and the quark-flavour sector of the Standard
Model. The main actor in these studies has been the B-meson system. In
2001, CP-violating effects were discovered and measured in the B-meson
system by the BaBar \cite{aubertet} and Belle \cite{abe} Colaboration.
A detailed investigation was also made of some benchmark, rare decay
modes such as $B^{o}_{d}\to J/\psi K_{s}$, $B^{o}_{d}\to \phi K_{s}$ and
$B^{o}_{d}\to \pi^{+}\pi^{-}$ and many others, for a recent review see R.
Fleischer \cite{fleischer}.  As of May 2006, it can be said that all
existing data on CP-violation and rare decays in the quark sector can
be described by the Standard Model within the theoretical and
experimental uncertainties. The recent discovery and measurement of
flavour conversion of solar \cite{altmann,smy,ahmad,aharmin}
atmospheric \cite{fukuda,ashie}, reactor \cite{bemporad,Araki} and
acelerator \cite{apollonio,aliu} neutrinos have conclusively
established that neutrinos have non-vanishing mass and they mix among
themselves much like the quarks, thereby providing the first evidence
of new physics beyond the Standard Model. The difference of the
squared neutrino masses and the mixing angles in the lepton mixing
matrix, $U_{PMNS}$, were determined, but neutrino oscillation data are
insensitive to the absolute value of neutrino masses and also to the
fundamental issue of wether neutrinos are Dirac or Majorana particles.
Upper bounds on neutrino masses were provided by the searches that
probe the neutrino mass values at rest: beta decay experiments
\cite{eitel}, neutrinoless double beta decay \cite{eliot} and precision
cosmology \cite{elgaroy}.

These recent experimental advances triggered an enormous theoretical
activity attempting to uncover the nature of this new physics. This
includes further developments of the already existing mechanisms and
theories such as GUT's and Supersymmetric Grand Unified Theories
(SUSY GUTs) and the appearence of new ideas and approaches implemented
at a variety of different energy scales.

Regardless of the energy scales at which those theoretical models are
built, the mechanisms for fermion mass generation and flavour mixing
can roughly be classified into four different types:
\begin{enumerate}
\item Texture zeroes,
\item Family or flavour symmetries
\item Radiative mechanisms and
\item Seesaw mechanisms
\end{enumerate}

These mechanisms are not disjoint but rather they are related and in
many cases they complement and support each other. In the last six or
seven years, important theoretical advances have been made in the
understanding of these four mechanisms. The following points should be
stressed.

1. Phenomenologically, some striking progress has been made with the
help of texture zeroes and flavour symmetries in specifying the
quantitative relationship between flavour mixing angles and quark or
lepton mass ratios \cite{fritzsch0,xing1} and
\cite{mondragon1,mondragon2}.

2. After all the recent developments, the seesaw mechanism with large
scale of the B-L violations still looks as the most appealing and
natural mechanism of neutrino mass generation. At the same time, it is
not excluded that some more complicated version of this mechanism is
realized \cite{mohapatra}.

3. Gran Unification plus supersymmetry in some form still looks like
the most plausible scenario of physics which naturally embeds the
seesaw mechanism.

4. At the same time, it seems now clear that the ``seesaw GUT''
scenario does not provide a complete understanding of the neutrino
masses and mixings as well as the quark masses and mixings or, in
other words, the flavour structure of the mass matrices. Some new
physics on top of this scenario seems essential. In this connection,
two important questions arise:
\begin{itemize}
\item the posible existence of new symmetries that show up mainly or
  only in the lepton sector
\item the need to understand the relation between quarks and leptons
  and the picture of flavour physics and CP violation in a unified
  way. The corresponding phenomenology is very rich.
\end{itemize}

These two issues point to the need of simpler models.

5. The search for simpler models starts by first constructing a low
energy theory with the Standard Model and a discrete non-Abelian
flavour symmetry group $\tilde{G}_{F}$ and then showing the possible
embeddings of this theory into a GUT, like $SO(10)$ or $SU(5)$.  The
discrete symmetry will therefore be a subgroup of $SO(3)_{f}$ or
$SU(3)_{f}$. Models in which the discrete non-Abelian flavour symmetry
is only broken at low energies became very popular in the last few
years
\cite{harari,derman,hall,ma0,kubo1,chen,grimus-la,frampton,aranda1,aranda2}
. The search for an adequate discrete group has concentrated on the
smallest subgroups of $SO(3)$ or $SU(3)$ that have at least one
singlet and one doublet irreducible representations to accomodate the
fermions in each family \cite{medeicos}.

To end this section, I will very briefly outline some recent
developments on these questions in which the participation of the
mexican community of particles and fields is visible.

\subsection{Flavour permutational symmetry and Fritzsch textures}
The non-Abelian flavour permutational symmetry $S_{3L}\otimes S_{3R}$ and
its explicit sequential breaking according to $S_{3L}\otimes S_{3R}\supset S_{3
  diag}\supset S_{2 diag}$ was used by A. Mondrag\'on and E.
Rodr\'{\i}guez-J\'auregui\cite{mondragon1,mondragon2} to characterize the
quark mass matrices, $M_{u}$ and $M_{d}$, with a texture of the same
modified Fritzsch type.  In a symmetry adapted basis, different
patterns for breaking the permutational symmetry give rise to quark
mass matrices which differ in the ratio $Z^{1/2} = M_{23}/M_{22}$ and
are labeled in terms of the irreducible representations of an
auxiliary $S_{2}$ group. After analytically diagonalizing the mass
matrices, these authors derive explicit, exact expressions for the
elements of the quark mixing matrix, $V_{CKM}^{th}$, the Jarlskog
invariant, $J$, and the three inner angles, $\alpha $, $\beta $ and $\gamma $ of the
unitarity triangle as functions of the quark mass ratios and only two
free parameters, the symmetry breaking parameter $Z^{1/2}$ and one
CP-violating phase $\Phi $.  The numerical values of these parameters
which characterize the experimentally preferred symmetry breaking
pattern $Z^{1/2} = 9/2\sqrt{2}$ and $\Phi = 90^{\circ }$, were extracted from
a $\chi^{2}$ fit of the theoretical expressions for the moduli,
$|V^{th}|$, to the experimentally determined values of the moduli of
the elements of the quark mixing matrix $|V^{exp}_{CKM}|$. The
agreement between theory and experiment, which initially was fairly
good, improved as the experimental determination of the elements of
the mixing matrix and the inner angles of the unitarity triangle
improved \cite{xing2}. The phase equivalence of $V^{th}_{CKM}$ and the
mixing matrix $V^{PDG}_{CKM}$ in the standard parametrization
advocated by the Particle Data Group allowed to translate those
results into explicit exact expressions for the three mixing angles
$\theta_{12}, \theta_{13}, \theta_{23}$ and the CP-violating phase $\delta_{13}$ in terms
of the four quark mass ratios and the symmetry breaking parameters
$Z^{1/2}$ and $\Phi$ \cite{mondragon2}.

The main point in these results is simply that the hierarchy of quark
masses and the texture of quark mass matrices are enough to determine,
at least partly, some important features of the quark flavour mixing.
In this sense, it was established that a scheme in which the two quark
mass matrices, $M_{u}$ and $M_{d}$, have the same modified Fritzsch
texture with the same value of the symmetry breaking parameter has
some predictive power for the flavour mixing angles and CP-violating phase.

\subsection{Models of flavour with continuous symmetry}
There is a large variety of possible candidates for supersymmetric
models of new physics beyond the Standard Model based in $N=1$ SUSY
with commuting GUT's and family symmetry groups, $G_{GUT}\otimes G_{f}$. This
is so because there are many possible candidate GUT's and family
symmetry groups $G_{f}$. The model dependence does not end there since
the details of the symmetry breaking vacuum plays a crucial role in
specifying the model and determining the masses and mixing angles, for
a recent review see S.F. King \cite{king0}. G. G. Ross and L.
Velasco-Sevilla chose the largest family symmetry group, $SU(3)$,
consistent with $SO(10)$ GUT's and with additional Abelian family
symmetries chosen to restrict the allowed Yukawa couplings. In a
series of interesting papers\cite{ross1,ross2,lileana} they explored
the phenomenological implications of their model and were able to find
a symmetry breaking scheme in which the observed hierarchical quark
masses and mixings are described together with the hierarchy of
charged lepton masses and a hierarchical structure for the neutrino
masses. The significant differences between quark and lepton mixings
are explained as due to the seesaw mechanism.  Given the very large
underlying symmetry, the fermion masses are heavily constrained. This
$SO(10)\otimes SU(3)_{f}$ model provides a consistent description of the
known masses and mixings of quarks and leptons. In the quark sector,
the presence of CP violating phases is necessary, not only to
reproduce CP violating processes, but also to reproduce the abserved
masses and mixings. In this model, the spontaneous breaking of CP in
the flavour sector naturally solves the supersymmetric CP problem and
the SUSY flavour problem, although flavour changing processes must
occur at a level close to current experimental bounds. Motivated by
the fact that leptogenesis is a very attractive candidate for
explaining the large baryon asymmetry observed in the universe L.
Velasco-Sevilla \cite{lileana} also explored the very interesting
possible connections between low energy CP violating phases appearing
in the lepton mixing matrix and those phases relevant for
leptogenesis.

\subsection{SUSYGUT Models with discrete flavour symmetry}
As noted above, in supersymmetric Grand Unified models of flavour with
a non-Abelian continuous family group, such as $SO(10)\otimes SU(3)_{f}$ or
$SU(5)\otimes SU(2)_{f}$, the phenomenological success depends crucially on
the details of the symmetry breaking vaccum and its alignement.

For instance, neutrino mixing angle relations such as the bimaximal
mixings of the left handed neutrinos is achieved only if the Yukawa
couplings involving different families are related in some special
way. The condition for the required equalities of Yukawa couplings to
emerge is that the several scalar fields which break the family
symmetry, called flavons, have their vacuum expectation values
carefully aligned (or misaligned) along special directions in family
space. Then, if these flavons appear in the effective operators
responsible for the Yukawa couplings, the relations between the Yukawa
couplings may be due to the particular alignement of the flavons
responsible for that particular operator. 

In an interesting series of papers, A. Aranda, C.D. Carone and R.F.
Lebed \cite{aranda3,aranda4,aranda5} showed that the physics of vacuum
alignement simplifies if the continuous family symmetry $SU(2)_{f}$ is
replaced by the discrete non-Abelian family symmetry $T'\otimes Z_{3}$ in
the SUSYGUT model of flavour $SU(5)\otimes SU(2)_{f}$ proposed by Romanino,
Barbieri and Hall \cite{barbieri1,barbieri2,barbieri3}. The group $T'$
is the group of proper rotations that leave a regular tetrahedron
invariant in the $SU(2)$ double covering of $SO(3)$.  It has singlet,
doublet and triplet irreducible representations with the
multiplication rule $2\otimes 2 = 1 \oplus 3$, which is a requisite to reproduce
the phenomenologically succesful mass textures derived from GUT
$SU(5)\otimes SU(2)_{f}.$ The extra Abelian $Z_{3}$ factor in $G_{f} = T'\otimes
Z_{3}$ is included in order to obtain the minimal extension needed to
reproduce the $SU(2)$ model textures and satisfy discrete anomaly
cancellation conditions. The flavons have non-trivial transformation
properties under the GUT SU(5) symmetry and the up-type and down-type
quark mass textures are accordingly modified. Additionally, in the
lepton sector, the rich representation structure of $T'$ allows for
the neutrinos to be placed in different reps than the charged leptons,
which, in this model is the origin of different hierarchies in the two
sectors. The symmetry breaking pattern is $T'\otimes Z_{3} \to
Z_{3}^{diag}\to\mbox{nothing}$. The light neutrino masses are generated
through the seesaw mechanism. Three generations of right-handed
neutrinos are introduced with the assignements $2^{0-}\oplus 1^{- +}$. This
assignement leads to Dirac and Majorana mass matrices that allow the
introduction of flavons that do not contribute at all to the charged
fermion mass matrices. In this way, mass matrices with a modified
Fritzsch texture are generated for the $u$ and $d-$type quarks, and
for the charged leptons while the light Majorana neutrino mass matrix
has a texture that naturally leads to the bimaximal $U_{PMNS}$ lepton
mixing matrix.

Some further advantages of using a finite, discrete family symmetry
are the following:
\begin{enumerate}
\item The breaking of a discrete symmetry does not lead to unwanted
  massless Goldstone bosons, unlike continuous symmetries
\item If this breaking is only spontaneous, it might produce domain
  walls \cite{zeldovich} which can be a serious problem.
  However, it can be solved by either invoking low scale inflation or
  embedding the discrete symmetry group into a continous group
  \cite{lazarides} as is the case for $T'\otimes Z_{3}\subset SU(2)$.
\item In the context of SUSY, discrete gauge symmetries do not give
  rise to excessive flavour changing neutral currents (FCNC) as is the
  case for continous symmetries.
\end{enumerate}

\subsection{Type II seesaw and $S_{3}\otimes U(1)_{e-\mu -\tau}$ symmetry} One of
the first phenomenologically succesful models for reproducing the
bimaximal mixing among the neutrinos was presented by R.N.  Mohapatra,
A. P\'erez-Lorenzana and C. Pires \cite{lorenzana}. This model is an
extension of the Standard Model where the bimaximal mixing pattern
among the neutrinos naturally arises via the type II seesaw mechanism.
The model does not include right handed neutrinos, the lepton content
of the SM is left unaltered but the Higgs sector is modified. The
$SU_{L}(2)$ content of the Higgs sector consists of three doublets,
two triplets with $Y= 2$ and a charged isosinglet with $Y = +2$. The
model has a global $S_{3}\otimes U(1)_{e-\mu -\tau}$ flavour symmetry. The
charged $\mu $ and $\tau$ fields are in doublet representations, while the
$e$ field is in a singlet representation of $S_{3}$. The pattern of
$SU_{L}(2)$ Higgs doublet vacuum expectation values leads to a
diagonal mass matrix for the charged leptons while the additional
Higgs triplet acquires naturally small vacuum expectation values due
to the type II see saw mechanism. At tree level, the $\nu_{\mu}$ and
$\nu_{e}$ masses are degenerate, but the presence of the global
$L_{e}-L_{\mu}-L_{\tau}$ and $S_{3}$ symmetry leads naturally to the
desired mass splitings among neutrinos at the one loop level. The
resulting neutrino masses have an inverted hierarchy $|m_{1}|\geq
|m_{2}|> > |m_{3}|$. There is a well known difficulty of this very
interesting model to fit the large angle solution of the solar
neutrino problem. Indeed, barring cancelletions between the
perturbations, these must be very small in order to obtain a
$\Delta m^{2}_{sun}$ close to the best fit value, but then, the value of
$\sin^{2}2\theta $ comes out too close to unity in disagreement with the best
global fits of solar data \cite{altarelli0}.

\subsection{A minimal $S_{3}-$invariant extension of the Standard Model}
The discovery of neutrino masses and mixings added ten new parameters
to the already long list of free parameters in the Standard Model and
made evident the urgent need of a systematic and unified treatment of
all fermions in the theory. These two facts, taken together, pointed
to the necessity and convenience of eliminating parameters and
systematizing the observed hierarchies of masses and mixings as well
as the presence or absence of CP violating phases by means of a
flavour or family symmetry under which the families transform in a
non-trivial fashion. As explained above, such a flavour symmetry
might be a continuous or, more economically, a finite group.

In a recent paper, J. Kubo, A. Mondrag\'on, M. Mondrag\'on and E.
Rodr\'{\i}guez-J\'auregui\cite{kubo1} argued that such a flavour
symmetry, unbroken at the Fermi scale, is the permutational symmetry
of three objects, $S_{3}$, and introduced a Minimal $S_{3}-$invariant
Extension of the Standard Model. In this model, $S_{3}$ is imposed as
a fundamental symmetry in the matter sector which is only
spontaneously broken together with the electroweak gauge symmetry.
This assumption leads to extend the concept of flavour and generations
to the Higgs sector. Hence, going to the irreducible representations
of $S_{3}$, the model has one Higgs $SU(2)_{L}$ doublet in the
$S_{3}-$singlet representation plus two more Higgs $SU(2)_{L}$
doublets which can only belong to the two components of the
$S_{3}-$doublet representation.  The fermion content of the Standard
Model is left unaltered. In this way, all the matter fields - Higgs,
quarks and lepton fields including the right-handed neutrino fields -
belong to the three dimensional representation $1_{s}\oplus 2$ of the
permutation group $S_{3}$. The leptonic sector is further constrained
by an Abelian $Z_{2}$ symmetry. A defined structure of the Yukawa
couplings is obtained which permits the calculation of mass and mixing
matrices for quarks and leptons in a unified way. The Majorana
neutrinos acquire mass via the type I seesaw mechanism. In a recent
paper, O. Felix, A. Mondrag\'on, M. Mondrag\'on and E. Peinado
\cite{ofelix} reparametrized the mass matrices of charged leptons and
neutrinos in terms of the respective mass eigenvalues and derived
explicit analytic and exact expressions in closed form for the mixing
angles appearing in the $U_{PMNS}$ matrix as functions of the masses
of charged leptons and neutrinos and one Majorana phase $\Phi_{\nu}$.  The
$U_{PMNS}$ matrix has also one Dirac phase which has its origin in the
charged lepton mass matrix. The numerical values of the mixing angles
$\theta_{13}$ and $\theta_{23}$ are determined by the mass of charged leptons
only in very good agreement with the best fit experimental values. The
solar mixing angle $\theta_{12}$ is almost insensitive to the values of the
masses of the charged leptons, but its experimental value allows the
determination of the neutrino mass spectrum which has an inverted
hierarchy with the values $|m_{\nu_{2}}| = $ 0.0507 eV, $|m_{\nu_{1}}| = $
0.0499 eV and $|m_{\nu_{3}}| = $ 0.0193 eV. A complete and detailed
discussion of the Majorana phases of the neutrino mixing matrix in
this model is given in J. Kubo \cite{kubo-u}.  A numerical analysis of
the quark mass matrices and the $V_{CKM}$ matrix gives one set of
parameters that are consistent with the experimental values given by
the Particle Data Group \cite{pdg}. A slightly less sketchy
explanation of this model is given in the next section.

\section{A minimal $S_{3}-$ invariant extension of the standard model}
Recently, a minimal $S_{3}-$invariant extension of the Standard Model
was suggested in \cite{kubo1}, in this section I will explain in a
slightly more detailed fashion the main ideas of this model and some
recent results on neutrino masses and mixings. 

{\bf $S_{3}-$symmetric Lagrangian and fermions masses .}
In the Standard Model analogous fermions in different generations have
completely identical couplings to all gauge bosons of the strong, weak
and electromagnetic interactions.  Prior to the introduction of the
Higgs boson and mass terms, the Lagrangian is chiral and invariant
with respect to permutations of the left and right fermionic fields.

The six possible permutations of three objects $(f_{1},f_{2},f_{3})$
are elements of the permutational group $S_{3}$. This is the discrete,
non-Abelian group with the smallest number of elements. The
three-dimensional real representation is not an irreducible
representation of $S_{3}$, it can be decomposed into the direct sum of
a doublet and a singlet, ${\bf 1}_{s} \oplus 2$. The direct product of two
doublets may be decomposed into the direct sum of two singlets and one
doublet, $ {\bf 2\otimes 2}={\bf 1}_{s}\oplus {\bf 1}_{A} + 2$. The antisymmetric
singlet is not invariant under $S_{3}$.

Since the Standard Model has only one Higgs $SU(2)_{L}$ doublet,
which can only be an $S_{3}$ singlet, it can only give mass to the
quark or charged lepton in the $S_{3}$ singlet representation, one in
each family, without breaking the $S_{3}$ symmetry.
Therefore, in order to impose $S_{3}$ as a fundamental symmetry,
unbroken at the Fermi scale, we are led to extend the concept of
flavour and generations to the Higgs sector of the theory. Hence,
going to the irreducible representations of $S_{3}$, we add to the
Higgs $SU(2)_{L}$ doublet in the $S_{3}-$ singlet representation, two
more $SU(2)_{L}$ doublet in the $S_{3}-$doublet representation. In
this way, all the quark, lepton and Higgs fields,
$Q^T=(u_L,d_L)~,~ u_R~,~d_R~,~L^T=(\nu_L,e_L)~,~e_R~,~ 
\nu_R~\mbox{ and }~H,$ are in reducible representations ${\bf 1}_{s}\oplus 2$. 
The most general  renormalizable Yukawa interactions are given by
\begin{eqnarray}\label{lag}
{\cal L}_Y = {\cal L}_{Y_D}+{\cal L}_{Y_U}
+{\cal L}_{Y_E}+{\cal L}_{Y_\nu},
\end{eqnarray}
where
\begin{eqnarray}\label{lagd}
{\cal L}_{Y_{D,E}} &=&
- Y_1^d \overline{ Q}_I H_S d_{IR} - Y_3^d \overline{ Q}_3 H_S d_{sR}  \cr
&  &   -Y^{d}_{2}[~ \overline{ Q}_{I}(\sigma_{1})_{IJ} H_1  d_{JR}
-\overline{ Q}_{I} (\sigma_{3})_{IJ} H_2  d_{JR}~]\cr
&  & -Y^d_{4} \overline{ Q}_s H_I  d_{IR} - Y^d_{5} \overline{ Q}_I H_I d_{sR} 
+~\mbox{h.c.} ,
\end{eqnarray}
\begin{eqnarray}\label{lagu}
{\cal L}_{Y_{U,\nu}} &=&
-Y^u_1 \overline{ Q}_{I}(i \sigma_2) H_S^* u_{IR} 
-Y^u_3\overline{ Q}_3(i \sigma_2) H_S^* u_{sR} \cr
&  &   -Y^{u}_{2}[~ \overline{ Q}_{I} (\sigma_{1})_{IJ} (i \sigma_2)H_1^*  u_{JR}
-  \overline{ Q}_{I} (\sigma_{3})_{IJ}(i \sigma_2) H_2^*  u_{JR}~]\cr
&  &
-Y^u_{4} \overline{ Q}_{s} (i \sigma_2)H_I^* u_{IR} 
-Y^u_{5}\overline{ Q}_I (i \sigma_2)H_I^*  u_{sR} +~\mbox{h.c.},
\end{eqnarray}
The fields in the $S_{3}-$doublets carry capital indices $I$ and $J$,
which run from 1 to 2 and the singlets are denoted by the subscript
$s$.

Furthermore, we add to the Lagrangian the Majorana mass terms for
the right-handed neutrinos 
\begin{eqnarray} 
{\cal L}_{M} = -M_1 \nu_{IR}^T C \nu_{IR} -M_3 \nu_{3R}^T C \nu_{3R}.
\label{majo}
\end{eqnarray}

Due to the presence of three Higgs fields, the Higgs potential
$V_H(H_S,H_D)$ is more complicated than that of the Standard Model.
This potential was analyzed by Pakvasa and Sugawara~\cite{pakvasa1},
see also Kubo\cite{kubo2}, who found that in addition to the $S_{3}$
symmetry, it has a permutational symmetry $S_{2}$: $H_{1}\leftrightarrow H_{2}$,
which is not a subgroup of the flavour group $S_{3}$ and an Abelian
discrete symmetry that will be used for selection rules of the Yukawa
couplings in the leptonic sector. Here, we will assume that the
vacuum respects the accidental $S_{2}$ symmetry of the Higgs potential
and $ \langle H_{1} \rangle = \langle H_{2} \rangle $.

With these assumptions, the Yukawa interactions, eqs. (\ref{lagd})-
(\ref{lagu}) yield mass matrices, for all fermions in the theory, of
the general form
\begin{equation}\label{general-m}
{\bf M} = \left( \begin{array}{ccc}
\mu_{1}+\mu_{2} & \mu_{2} & \mu_{5} \\  
\mu_{2} & \mu_{1}-\mu_{2} &\mu_{5}\\
\mu_{4} & \mu_{4}&  \mu_{3}
\end{array}\right).
\end{equation}

The Majorana masses for the left neutrinos $\nu_{L}$ will be obtained
from the see-saw mechanism. The corresponding mass matrix is
given by
\begin{eqnarray}
{\bf M_{\nu}} = {\bf M_{\nu_D}}\tilde{{\bf M}}^{-1}({\bf M_{\nu_D}})^T
\label{seesaw}
\end{eqnarray}
where $\tilde{{\bf M}}=\mbox{diag}(M_1,M_1,M_3)$.

In principle, all entries in the mass matrices can be complex since
there is no restriction coming from the flavour symmetry $S_{3}$.

The mass matrices are diagonalized by bi-unitary transformations as
\begin{equation}
\begin{array}{rcl}
U_{d(u,e)L}^{\dag}{\bf M}_{d(u,e)}U_{d(u,e)R} 
&=&\mbox{diag} (m_{d(u,e)}, m_{s(c,\mu)},m_{b(t,\tau)}),\\ 
U_{\nu}^{T}{\bf M_\nu}U_{\nu} &=&
\mbox{diag} (m_{\nu_1},m_{\nu_2},m_{\nu_3}).
\end{array}
\label{unu}
\end{equation}
The entries in the diagonal matrices may be complex, so the physical
masses are their absolute values.

The mixing matrices are, by definition,
\begin{equation}
\begin{array}{ll}
V_{CKM} = U_{uL}^{\dag} U_{dL},& V_{PMNS} = U_{eL}^{\dag} U_{\nu}.
\label{ckm1}
\end{array}
\end{equation}
\subsection{Leptonic sector and $Z_{2}$ symmetry.} 
A further reduction of the number of parameters in the
leptonic sector may be achieved by means of an Abelian $Z_{2}$
symmetry. A set of charge assignments of $Z_{2}$, compatible with the
experimental data on masses and mixings in the leptonic sector is
given in Table 1
\begin{center}
\begin{tabular}{|c|c|}
\hline
 $-$ &  $+$
\\ \hline

$H_S, ~\nu_{3R}$ & $H_I, ~L_3, ~L_I, ~e_{3R},~ e_{IR},~\nu_{IR}$
\\ \hline
\end{tabular}
\end{center}
\vspace{-0.3cm}
\begin{center}
{\footnotesize {\bf Table I}. $Z_2$ assignment in the leptonic sector.}
\end{center}
These $Z_2$ assignments forbid certain Yukawa couplings,
\begin{eqnarray}
 Y^e_{1} = Y^e_{3}= Y^{\nu}_{1}= Y^{\nu}_{5}=0.
\label{zeros}
\end{eqnarray}

Therefore, the corresponding entries in the mass matrices vanish,{\it i.e.},
$\mu_{1}^{e}=\mu_{3}^{e}=0$ and $\mu_{1}^{\nu}=\mu_{5}^{\nu}=0$.
\subsection{ The mass matrix of the charged leptons}
The mass matrix of the charged leptons takes the form
\begin{equation}
M_{e} = m_{\tau}\left( \begin{array}{ccc}
\tilde{\mu}_{2} & \tilde{\mu}_{2} & \tilde{\mu}_{5} \\  
\tilde{\mu}_{2} &-\tilde{\mu}_{2} &\tilde{\mu}_{5}\\ 
\tilde{\mu}_{4} & \tilde{\mu}_{4}& 0
\end{array}\right).
\label{charged-leptons-m}
\end{equation}
The unitary matrix $U_{eL}$ that enters in the definition of the
mixing matrix, $U_{PMNS}$, is calculated from
\begin{eqnarray}\label{unita}
U_{eL}^{\dag}M_{e}M_{e}^{\dag}U_{eL}=\mbox{diag}(m_{e}^{2},m_{\mu}^{2},m_{\tau}^{2}),
\end{eqnarray}
The entries in the mass matrix squared, $M_{e}M^{\dagger}_{e}$, may readily
be expressed in terms of the mass eigenvalues $(m^{2}_{e}, m^{2}_{\mu},
m^{2}_{\tau})$.  Then, the matrix $U_{eL}$ may be expressed in terms of
the charged lepton masses and one Dirac phase,
\begin{equation}
U_{eL}\approx \left(
\begin{array}{ccc} 
\frac{1}{\sqrt{2}}\frac{\frac{m_{e}}{m_{\mu}}}{\sqrt{1-\left(\frac{m_{e}}{m_{\mu}}
\right)^{2}}}
&\frac{1}{\sqrt{2}}\frac{1}{\sqrt{1+\left(\frac{m_{e}}{m_{\mu}}\right)^{2}}}
&\frac{1}{\sqrt{2}}\frac{e^{i\delta_{e}}}{\sqrt{1+\frac{m_{e}m_{\mu}}{m_{\tau}^{2}}}}
\\
-\frac{1}{\sqrt{2}}\frac{\frac{m_{e}}{m_{\mu}}}{\sqrt{1-\left(\frac{m_{e}}{m_{\mu}}
\right)^{2}}}
&-\frac{1}{\sqrt{2}}\frac{1}{\sqrt{1+\left(\frac{m_{e}}{m_{\mu}}\right)^{2}}}  
& \frac{1}{\sqrt{2}}\frac{e^{i\delta_{e}}}{\sqrt{1+\frac{m_{e}m_{\mu}}{m_{\tau}^{2}}}}
\\ 
\frac{\sqrt{1-2\left(\frac{m_{e}}{m_{\mu}}\right)^{2}}}{\sqrt{1-
\left(\frac{m_{e}}{m_{\mu}}\right)^{2}}}&\frac{\frac{m_{e}}{m_{\mu}}}
{\sqrt{1+\left(\frac{m_{e}}{m_{\mu}}\right)^{2}}}  &\frac{\frac{m_{e}m_{\mu}}
{m_{\tau}^{2}}e^{i\delta_{e}} }{\sqrt{1+\frac{m_{e}m_{\mu}}{m_{\tau}^{2}}}}
\end{array}\right) 
\end{equation}
\subsection{ The mass matrix of the neutrinos}
According with the $Z_{2}$ selection rule eq. (\ref{zeros}),$\mu^{\nu_{D}}_{1}=
\mu^{\nu}_{5} = 0$ in (\ref{charged-leptons-m}).
Then, the mass matrix for the left-handed Majorana neutrinos obtained
from the see-saw mechanism takes the form
\begin{eqnarray}\label{m-nu}
{\bf M_{\nu}} &=& {\bf M_{\nu_D}}\tilde{{\bf M}}^{-1} 
({\bf M_{\nu_D}})^T \cr 
&=& \pmatrix{
m_{\nu_{3}} & 0 & \sqrt{(m_{\nu_{3}}-m_{\nu_{1}})(m_{\nu_{2}}-m_{\nu_{3}})}\cr 
0 &m_{\nu_{3}}  & 0 \cr
\sqrt{(m_{\nu_{3}}-m_{\nu_{1}})(m_{\nu_{2}}-m_{\nu_{3}})}  & 0  & m_{\nu_{1}}+
m_{\nu_{2}}-m_{\nu_{3}}
} 
\end{eqnarray}
as in the case of the charged leptons, the matrix $M_{\nu_{D}}$ has been
reparametrized in terms of its eigenvalues, the complex neutrino
masses.

The unitary matrix $U_{\nu}$ that brings $M_{\nu_{D}}$ to diagonal form is
\begin{equation}
U_{\nu}=\left(
\begin{array}{ccc}
\sqrt{\displaystyle{\frac{m_{\nu_{2}}-m_{\nu_{3}}}
{m_{\nu_{2}}-m_{\nu_{1}}}}}& \sqrt{
\displaystyle{\frac{m_{\nu_{3}}-m_{\nu_{1}}}{m_{\nu_{2}}-m_{\nu_{1}}}}} & 0\\
0&0&1\\
-\sqrt{\displaystyle{\frac{m_{\nu_{3}}-m_{\nu_{1}}}{m_{\nu_{2}}-m_{\nu_{1}}}}}  &
\sqrt{\displaystyle{\frac{m_{\nu_{2}}-m_{\nu_{3}}}
{m_{\nu_{2}}-m_{\nu_{1}}}}}&0
\end{array}
\right).
\end{equation}
The unitarity of $U_{\nu}$ constrains its entries to be real. This
condition fixes the phases $\phi_{1}$ and $\phi_{2}$ as 
\begin{eqnarray}\label{rela}
|m_{\nu_{1}}|\sin\phi_{1}=|m_{\nu_{2}}|\sin\phi_{2}=|m_{\nu_{3}}|\sin\phi_{\nu} = 0
\end{eqnarray}
The only free parameter in $M_{\nu}$ and $U_{\nu}$, other than the real
neutrino masses $|m_{\nu_{1}}|$, $|m_{\nu_{2}}|$ and $|m_{\nu_{3}}|$, is the
phase $\phi_{\nu}$.
\subsection{The neutrino mixing matrix}

The neutrino mixing matrix $V_{PMNS}$, in the standard form advocated
by the $PDG$, is obtained by taking the product $U_{eL}^{\dag}U_{\nu}$ and
making an appropriate transformation of phases, $U_{PMNS}$ is, then
equal to
\begin{equation}\label{vpmns}
\left( \begin{array}{ccc}
\frac{1}{\sqrt{2}}\frac{x}{\sqrt{1-x^2}}\sin \eta +\frac{\sqrt{1-2x^2}}
{\sqrt{1-x^2}}\cos \eta  & \frac{1}{\sqrt{2}}\frac{x}{\sqrt{1-x^2}}
\cos \eta -\frac{\sqrt{1-2x^2}}{\sqrt{1-x^2}}\sin \eta  & -\frac{1}{\sqrt{2}}
\frac{x}{\sqrt{1-x^2}}e^{-i\delta_{e}}\\
\\
\frac{1}{\sqrt{2}}\frac{1}{\sqrt{1+x^2}}\sin \eta  - \frac{x}{\sqrt{1-x^2}}
\cos \eta e^{i\delta_{e}}  &\frac{1}{\sqrt{2}}\frac{1}{\sqrt{1+x^2}}\cos \eta +
\frac{x}{\sqrt{1-x^2}}\sin \eta e^{i\delta_{e}}  & -\frac{1}{\sqrt{2}}
\frac{1+2\frac{z}{y(1-y)}}{\sqrt{1+x^2}} \\
\\
\frac{1}{\sqrt{2}}\frac{1}{\sqrt{1+\sqrt{z}}}\sin \eta -\frac{\sqrt{z}}
{\sqrt{1+\sqrt{z}}}\cos \eta e^{i\delta_{e}}  & \frac{1}{\sqrt{2}}\frac{1}
{\sqrt{1+\sqrt{z}}}\cos \eta    +\frac{\sqrt{z}}{\sqrt{1+\sqrt{z}}}
\sin \eta e^{i\delta_{e}}  & \frac{1}{\sqrt{2}}\frac{1}{\sqrt{1+\sqrt{z}}}
\end{array}\right)K,
\end{equation}
where 
\begin{eqnarray}
  \sin\eta = \sqrt{\frac{m_{\nu_{2}}-m_{\nu_{3}}}{m_{\nu_{2}}-m_{\nu_{1}}} } 
  \hspace{0.5cm}  x = \frac{m_{e}}{m_{\mu}}, 
  \hspace{0.4cm} y = \frac{m^{2}_{e}+m^{2}_{\mu}}{m^{2}_{\tau}},\hspace{0.4cm} 
\mbox{and} \hspace{0.4cm}
  z= \Bigl(\frac{m_{e}m_{\mu}}{m^{2}_{\tau}}\Bigr)^{2}
\end{eqnarray}
and $K = \mbox{diag}(1,e^{i\alpha},e^{i\beta})$ is the diagonal matrix of the
Majorana phases.

Explicit expressions for the mixing angles in terms of the lepton
masses are obtained from a comparison of $U^{th}_{PMNS}$, eq.(\ref{vpmns}) with
the standard parametrization advocated by the PDG\cite{pdg}.
\begin{eqnarray}
\tan\theta_{12}\approx\sqrt{\frac{( |m_{\nu_{2}}|^{2}-|m_{\nu_{3}}|^{2}\sin^{2}\phi_{\nu})^{1/2}-
|m_{\nu_{3}}\|cos\phi_{\nu}|}{( |m_{\nu_{2}}|^{2}-|m_{\nu_{3}}|^{2}\sin^{2}\phi_{\nu})^{1/2}+
|m_{\nu_{3}}\|\cos\phi_{\nu}|} },
\end{eqnarray}
\begin{eqnarray}\label{se}
\sin \theta_{13}\approx
\frac{\frac{1}{\sqrt{2}}\frac{m_{e}}{m_{\mu}}}{\sqrt{1-\left(\frac{m_{e}}
{m_{\mu}}\right)^2}},\hspace{0.5cm}\mbox{and}\hspace{0.5cm}
\sin \theta_{23}\approx
-\frac{1}{\sqrt{2}}\frac{\sqrt{1-\left(\frac{m_{e}}{m_{\mu}}\right)^2}}
{\sqrt{1-\frac{1}{2}\left(\frac{m_{e}}{m_{\mu}}\right)^2}}
\end{eqnarray}
Similarly, the Majorana phases are given by
\begin{equation}
\begin{array}{l}
\sin 2\alpha=\sin(\phi_{1}-\phi_{2})=
\frac{|m_{\nu_{3}}|\sin\phi_{\nu}}{|m_{\nu_{1}}||m_{\nu_{2}}|}\times\\
\left(\sqrt{|m_{\nu_{2}}|^2-|m_{\nu_{3}}|^{2}\sin^{2}\phi_{\nu}}+
\sqrt{|m_{\nu_{1}}|^{2}-|m_{\nu_{3}}|^{2}\sin^{2}\phi_{\nu}}\right)
\\
\sin 2\beta=\sin(\phi_{1}-\phi_{\nu})=\\
 \frac{\sin\phi_{\nu}}{|m_{\nu_{1}}|}\left(|m_{\nu_{3}}|\sqrt{1-\sin^{2}\phi_{\nu}}+
\sqrt{|m_{\nu_{1}}|^{2}-|m_{\nu_{3}}|^{2}\sin^{2}\phi_{\nu}}\right)
\end{array}
\end{equation}
A detailed discussion of the Majorana phases in the
neutrino mixing matrix $U_{PMNS}$ obtained in our model is given in 
J. Kubo~\cite{kubo-u}.
\subsection{Neutrino masses and mixings}
In this model, $\sin^{2} \theta_{13}$ and $\sin^{2} \theta_{23}$ are determined
by the masses of the charged leptons in very good agreement with the
experimental values~\cite{jung,fogli1,Maltoni:2004ei,schwetz},
\begin{equation}
\begin{array}{ll}
(\sin^{2}\theta_{13})^{th}=1.1\times 10^{-5}, &(\sin^2
  \theta_{13})^{exp} \leq 0.046, \nonumber
\end{array}
\end{equation}
and
\begin{equation}
\begin{array}{ll}
(\sin^{2}\theta_{23})^{th}=0.49, &(\sin^2
  \theta_{23})^{exp}=0.5^{+0.06}_{-0.05}.\nonumber
\end{array}
\end{equation}
In the present model, the experimental restriction $|\Delta
m^2_{12}|<|\Delta m^2_{13}|$ implies an inverted neutrino mass
spectrum, $|m_{\nu_{3}}|<|m_{\nu_{1}}|<|m_{\nu_{2}}|$~\cite{kubo1}.

The mass $|m_{\nu_{2}}|$ assumes its minimal value when $\sin
\phi_{\nu}$ vanishes, then
\begin{eqnarray}
|m_{\nu_{2}}|\approx \frac{\sqrt{\Delta
  m^2_{13}}}{\sin 2\theta_{12}}.
\end{eqnarray}
Hence, we find
\begin{eqnarray}
|m_{\nu_{2}}|\approx0.0507eV,\hspace{0.3cm} |m_{\nu_{1}}|\approx 0.0499eV, \hspace{0.3cm}
|m_{\nu_{3}}|\approx 0.0193eV
\end{eqnarray}
where we used the values $\Delta m^{2}_{13}=2.2^{+0.37}_{-0.27}\times
10^{-3}eV^{2}$ and $\sin^{2}\theta_{12}=0.31^{+0.02}_{-0.03}$ taken
from M. Maltoni et al.~\cite{Maltoni:2004ei,schwetz} and
G. L. Fogli et al.~\cite{fogli1}.

With those values for the neutrino masses we compute the effective
electron neutrino mass $m_{\beta}$,
\begin{equation}\label{mbet}
 \label{mb} m_\beta = \left[\sum_i|U_{ei}|^2m^2_{\nu_{i}}\right]^\frac{1}{2}=0.0502eV ,
\end{equation}
well below the upper bound $m_{\beta}<1.8eV$ coming from the tritium
$\beta$-decay experiments~\cite{eitel,fogli1,fogli2}.

\subsection{The hadronic sector}
The $Z_{2}$ assignements in the hadronic sector are independent of
those in the leptonic sector. Hence, in principle, it can be assumed
that $Z_{2}$ is a good symmetry at a more fundamental level and verify
that $Z_{2}$ is free from any quantum anomaly. However, if we give
all quarks even parity as we did to the charged leptons, the Yukawa
couplings $Y^{u,d}_{1}$ and $Y^{u,d}_{3}$ will be forbidden.
Consequently, the squared mass matrix, of the $u-$quarks,
$M_{u}M^{\dagger}_{u}$ would have a texture similar to the texture of the mass
matrix of the Majorana neutrinos, given in (\ref{charged-leptons-m}),
which would lead to a $U_{u}$ that produces large values of the quark
mixing angles in disagreement with the small experimental values.

Therefore, to give one set of parameters that are consistent with the
experimental values given by the Particle Data Group\cite{pdg}, and show
that the model is phenomenologically viable, we proceeded under the
assumption that $Z_{2}$ is explicitly broken in the hadronic sector.
Since all the $S_{3}$ invariant Yukawa couplings are now allowed, the
mass matrices for the quarks take the general form (\ref{general-m}),
where all the entries can be complex.  One can easily see that all the
phases, except for those of $\mu_1^{u,d}$ and $\mu_3^{u,d}$, can be
removed through an appropriate redefinition of the quark fields. Of
course, only one of the four phases of $\mu_1^{u,d}$ and $\mu_3^{u,d}$ is
observable in $V_{\rm CKM}$.  So, we assume that only $m_3^{d}$ is a
complex number. 

The gross structure of realistic mass matrices can be obtained, if
$\mu_3^{u,d} \sim O(m_{t,b})$ and $\mu_{1,2}^{u,d} \sim O(m_{c,s})$ (to achieve
realistic mass hierarchies), and the non-diagonal elements
$\mu_{4}^{u,d}$ and $\mu_{5}^{u,d} $ along with $\mu_{1,2}^{u,d}$ can
produce a realistic mixing among the quarks.  There are $10$ real
parameters and one phase to produce six quark masses, three mixing
angles and one CP-violating phase. The set of dimensionless parameters
\begin{eqnarray}
m_1^u/m_0^u &=& -0.000293~,~ m_2^u/m_0^u =-0.00028 ~,~   m_3^u/m_0^u=1~,\cr  
m_4^u/m_0^u &=&0.031  ~,~m_5^u/m_0^u=0.0386,\cr
m_1^d/m_0^d &=&0.0004~,~  m_2^d/m_0^d =0.00275~,~   m_3^d/m_0^d=1+ 1.2I~,\cr   
m_4^d/m_0^d  &=& 0.283~,~  m_5^d /m_0^d =0.058
\label{choice1}
\end{eqnarray}
yields the mass hierarchies
\begin{eqnarray}\label{ra}
m_u/m_t &=& 1.33\times 10^{-5}~,~m_c/m_t=2.99\times 10^{-3},\cr
m_d/m_b &=& 1.31\times 10^{-3}~,~m_s/m_b=1.17\times 10^{-2},
\end{eqnarray}
where $m_0^u=\mu_3^u$ and $m_0^d=\mbox{Re}(\mu_3^d)$, and
the mixing matrix becomes
\begin{eqnarray}\label{vckm1}
V_{\rm CKM} &=& U_{uL}^{\dag} U_{dL}\cr
&=&\pmatrix{
0.968+0.117 I & 0.198+0.0974 I &-0.00253-0.00354 I\cr
  -0.198+0.0969 I & 0.968-0.115 I  & -0.0222-0.0376 I  \cr
 0.00211+0.00648 I &0.0179-0.0395 I & 0.999-0.00206 I}.
\end{eqnarray}
The magnitudes of the elements are given by
\begin{eqnarray}\label{vckm2}
|V_{\rm CKM}| &=&
\pmatrix{
0.975& 0.221 &0.00435 \cr
0.221 & 0.974  & 0.0437\cr
0.00682 &0.0434 & 0.999},
\end{eqnarray}
which should be compared with the experimental values\cite{pdg}
\begin{eqnarray}
|V_{\rm CKM}^{\rm exp}| &=&
\pmatrix{
0.9741~\mbox{to}~ 0.9756 & 0.219 ~\mbox{to}~ 0.226 & 0.0025 
~\mbox{to}~ 0.0048\cr  
0.219 ~\mbox{to}~ 0.226  & 0.9732 ~\mbox{to}~ 0.9748   & 0.038 
~\mbox{to}~ 0.044\cr
 0.004 ~\mbox{to}~ 0.014  &0.037 ~\mbox{to}~ 0.044  & 0.9990 
~\mbox{to}~ 0.9993}.
\end{eqnarray}
Note that the mixing matrix (\ref{vckm1}) is NOT in the standard
parametrization.  So, we give the invariant measure of
CP-violations\cite{jarskog} 
\begin{eqnarray}\label{jar} 
J &=&\mbox{Im}~[(V_{\rm
  CKM})_{11}(V_{\rm CKM})_{22} (V_{\rm CKM}^*)_{12}(V_{\rm
  CKM}^*)_{21}]=2.5\times 10^{-5} 
\end{eqnarray} 
for the choice (\ref{choice1}), which is slightly larger than the
experimental value $(3.0\pm 0.3)\times 10^{-5}$ (see \cite{pdg} and also
\cite{b-factory}).  The angles of the unitarity triangle for $V_{CKM}$
(\ref{vckm1}) are given by
\begin{eqnarray}\label{an}
\phi_1 \simeq 22^\circ~,~ \phi_3 \simeq 38^\circ, 
\end{eqnarray} 
where the experimental values are: $\phi_1 = 24^\circ\pm 4^\circ$ and $\phi_3 = 59^\circ \pm
13^\circ$ \cite{pdg}.  The normalization masses $m_0^u$ and $m_0^d$ are
fixed at
\begin{eqnarray} 
m_0^u =174 ~~\mbox{GeV}
~,~m_0^d=1.8~~\mbox{GeV} 
\end{eqnarray} 
for $m_t=174$ GeV and $m_b=3$ GeV, yielding that $m_u \simeq 2.3 $ MeV,
$m_c\simeq 0.52$ GeV, $m_d \simeq 3.9 $ MeV and $ m_s= 0.035 $ GeV.  Although
these values cannot be directly compared with the running masses,
because our calculation is of the tree level, it is nevertheless
worthwhile to observe how close they are to \cite{fritzsch0}
\begin{eqnarray}\label{maqu} 
m_u(M_Z) &=& 0.9 -2.9~~\mbox{MeV}~,
~m_d(M_Z) = 1.8 -5.3~~\mbox{MeV},\cr
~m_c(M_Z) &=& 0.53 -0.68~~\mbox{GeV}~,
~m_s(M_Z) = 0.035 -0.100~~\mbox{GeV},\cr
~m_t(M_Z) &=& 168 -180~~\mbox{GeV}~, ~m_b(M_Z) = 2.8 -3.0~~\mbox{GeV}.
\end{eqnarray}

\section{conclusions}
The recent advances in the experimental knowledge of flavour physics,
CP-violation and fermion masses and mixings triggered an enormuous
theoretical activity aimed to uncover the nature of this new physics.
Important advances have been made in the further development of the
already existing mechanisms and theories and the proposal of ingenious
new ideas.

As an instance of the first approach, here I discussed the unified
SUSY SO(10) theory with an additional $SU(3)_{f}$ flavour symmetry
explored by G.G. Ross and L. Velasco-Sevilla
\cite{ross1,ross2,lileana}. The phenomenological success of this kind
of theories with a continuous gauged family symmetry is achieved
through the details of the symmetry breaking vacuum and elaborate
mechanisms for its alignement.  In this class of models, the physics
of vacuum alignement simplifies if the continuum family symmetry is
replaced by a discrete non-Abelian family symmetry as shown by A.
Aranda, C.D. Carone and R.F. Lebed \cite{aranda3,aranda4,aranda5} who
replaced the discrete non-Abelian group $T'\otimes Z_{3}$ for $SU(2)_{f}$ in
the SUSY $SU(5)\times SU(2)$ unified theory of Barbieri, Romanino et al
\cite{barbieri1,barbieri2,barbieri3}, and found that the reduction of
the underlying continuous family symmetry to a discrete subgroup
renders the desired vacuum alignement a generic property of such
models.

As an example of the second approach, I discussed two extensions of
the Standard Model in which the Higgs sector is modified and have an
additional $S_{3}$ non-Abelian symmetry. They have this symmetry in
common with the phenomenologically succesful efforts of A. Mondrag\'on
and E. Rodr\'{\i}guez-J\'auregui \cite{mondragon1,mondragon2} to uncover
a flavour $S_{3}$ symmetry in the Fritzsch texture zeroes of the quark mass
matrices and the $V_{CKM}$ phenomenology. 

In the model proposed and discussed by R. N. Mohapatra, A.
P\'erez-Lorenzana and C.A. de S. Pires \cite{lorenzana}, there are no right
handed neutrinos but additional Higgs triplets which acquire naturally
small vacuum expectation values due to the type II see-saw mechanism.
The presence of a global $S_{3}\otimes U(1)_{e-\mu-\tau}$ symmetry leads
naturally to the desired neutrino mass textures and generates the
desired small splittings among neutrinos in fair agreement with
experiment. 

In the Minimal $S_{3}-$invariant Extension of the Standard Model
proposed by J. Kubo, A. Mondrag\'on, M.  Mondrag\'on and E.
Rodr\'{\i}guez-J\'auregui \cite{kubo1}, the concept of flavour and
generations is extended to the Higgs sector by introducing three Higgs
fields that are $SU(2)_{L}$ doublets in such away that all matter
fields - lepton, quark and Higgs fields - belong to the three
dimensional reducible $1\oplus 2$ representation of the permutation group
$S_{3}$. A well defined structure of the Yukawa couplings is obtained
which permits the calculation of mass and mixing matrices for quarks
and leptons in a unfied way. A further reduction of redundant
parameters is achieved in the leptonic sector by introducing a $Z_{2}$
symmetry. In this model, the Majorana neutrinos acquire mass via the
type I see-saw mechanism. The flavour symmetry group $S_{3}\otimes
Z_{2}$ relates the mass spectrum and mixings. This allows the
computation of the neutrino mixing matrix explicitly in terms of the
masses of the charged leptons and neutrinos \cite{ofelix}. The
magnitudes of the three neutrino mixing angles are determined by the
interplay of the flavour $S_{3}\times Z_{2}$ symmetry, the see-saw
mechanism and the charged lepton mass hierarchy.  It is also found
that the lepton mixing matrix $V_{PMNS}$ has one Dirac CP-violating
phase and two Majorana phases. The numerical values of the $\theta_{13}$
and $\theta_{23}$ mixing angles are determined by the charged leptons only
in very good agreement with experiment. The solar mixing angle
$\theta_{12}$ is almost insensitive to the values of the masses of the
charged leptons but its experimental value allows to fix the scale and
origin of the neutrino mass spectrum which has an inverted hierarchy
with the values $|m_{\nu_{2}}| = $0.0507 eV $|m_{\nu_{1}}|= $ 0.0499 eV
and $|m_{\nu_{3}}| = $0.0193 eV.

In conclusion, a discernible trend is perceptible in the formulation
of symmetry based models of flavour,fermion masses and mixings and CP
violation. In a bottom up approach, the search for simpler models
starts with the formulation of a phenomenologically succesful low energy
theory with a minimal extension of the Standard Model and a discrete,
non-Abelian flavour or family group $\tilde{G}_{f}$, and then, showing the
possible embeddings of this theory into an SO(10) or SU(5) GUT and a
continuous flavour group $\tilde{G}_{f}\subset G_{f}$.


\begin{theacknowledgments}
This work was partially supported by CONACYT M\'exico under contract
No 40162-F and by DGAPA-UNAM under contract PAPIIT-IN116202-3.  
\end{theacknowledgments}



\end{document}